\documentclass{article}

\usepackage{multicol}
\usepackage{graphicx}
\usepackage{color}
\usepackage{amsmath}
\usepackage{amssymb}
\usepackage{enumitem}
\usepackage[margin=0.6in]{geometry}
\usepackage{MnSymbol}
\usepackage{cite}
\usepackage{authblk}
\usepackage[hidelinks]{hyperref}
\usepackage[version=4]{mhchem}
\begin{document}

\title{On the performance of optical phased array technology for beam steering}

\author[1,2,3,4,$^*$]{Antonio Cal\`a Lesina}
\author[5]{Dominic Goodwill}
\author[5]{Eric Bernier}
\author[1,2]{Lora Ramunno}
\author[1,2,3]{Pierre Berini}

\affil[1]{Department of Physics, University of Ottawa, Ottawa, Canada.}
\affil[2]{Centre for Research in Photonics, University of Ottawa, Ottawa, Canada.}
\affil[3]{School of Electrical Engineering and Computer Science, University of Ottawa, Ottawa, Canada.}
\affil[4]{Hannover Centre for Optical Technologies and Cluster of Excellence PhoenixD, Leibniz University Hannover, Hannover, Germany.}
\affil[5]{Huawei Technologies Canada, Ottawa, Canada.}
\affil[$^*$]{antonio.calalesina@hot.uni-hannover.de}

\maketitle
\begin{abstract} 
\noindent 
Optical phased arrays are of strong interest for beam steering in telecom and LIDAR applications. A phased array ideally requires that the field produced by each element in the array (a pixel) is fully controllable in phase and amplitude (ideally constant).
This is needed to realize a phase gradient along a direction in the array, and thus beam steering in that direction.
In practice, grating lobes appear if the pixel size is not sub-wavelength, which is an issue for many optical technologies. Furthermore, the phase performance of an optical pixel may not span the required $2\pi$ phase range, or may not produce a constant amplitude over its phase range. These limitations result in imperfections in the phase gradient, which in turn introduce undesirable secondary lobes.
We discuss the effects of non-ideal pixels on beam formation, in a general and technology-agnostic manner. 
By examining the strength of secondary lobes with respect to the main lobe, we quantify beam steering quality, and make recommendations on the pixel performance required for beam steering within prescribed specifications.
By applying appropriate compensation strategies, we show that it is possible to realize high-quality beam steering even when the pixel performance is non-ideal, with intensity of the secondary lobes be two orders of magnitude smaller than the main lobe. 

\end{abstract}


\section{Optical phased array technologies}

Optical phased arrays (OPAs) are of strong interest for applications requiring random-access pointing, multiple beam forming, and dynamic beam generation and steering, such as light detection and ranging (LIDAR) technology for autonomous machines, self-driving cars, smart antennas, and inter-satellite communications \cite{Heck2017,Sun2019,Fu2020,Shaltout2019}.
LIDARs require a beam steering function to enable ranging over a scan area of interest. Some commercial LIDARs implement this function by mechanic rotation \cite{Velodyne}. However, mechanic systems are bulky, not easy to integrate without compromising the aesthetics, and slow for self-driving vehicles. 
An OPA is a 2D array of pixels, and the steering function is implemented by creating a phase gradient over the surface via electronic control. Pixels can be coherent light emitters or scatterers. 

Pixels based on emitters may be implemented, {\it e.g}, as vertical-cavity surface-emitting lasers (VCSELs) \cite{Xun2016,Pan2019a,Sayyah2015}, while pixels based on scatterers modify the properties of light upon reflection or transmission.
Conventionally, a reflective phased array may be implemented in liquid crystal on silicon (LCoS) \cite{Zhang2014} technology (often termed spatial light modulators), where each pixel reflects the incident light with a variable phase. However, the requirement to re-orient molecules in a liquid by applying an electric field makes LCoS technology slow (response time of the order of milliseconds) \cite{Resler1996,McManamon1996,McManamon2009,Wu2018}. Solutions based on microelectromechanical systems (MEMS) mirrors have also made much progress allowing microsecond response time \cite{Chan2013,Yang2014a,Yoo2014,Holsteen2019,Wang2019}. Pixels in LCoS and MEMS can provide high reflectance and $2\pi$ phase range. However, they are large, thus leading to grating lobes and a limited angular range of operation, also known as field-of-view (FoV).

Silicon photonics and photonic integrated circuits (PIC) offer a promising platform for OPAs by allowing full electronic control and high tuning speed, which is required in LIDAR for real-time applications \cite{Sun2019,Sun2013,Heck2017,Abediasl2015,Zhao2017b}. 
Waveguide gratings in silicon-on-insulator (SOI) technology have been extensively studied for 1D \cite{Hulme2015,Xu2019} and 2D \cite{Kwong2014a,VanAcoleyen2009,VanAcoleyen2010,Doylend2011,Doylend2012,Yaacobi2014,Poulton2017} beam steering. In waveguide arrays, 2D beam steering is generally realized by exploiting two different physical principles: phase tuning of the waveguides for steering along one axis, and wavelength tuning of a laser source in conjunction with a grating, for steering along the other axis \cite{VanAcoleyen2009,Doylend2011,Xie2019,Miller2020}.
In waveguide arrays, the pitch is typically large to avoid crosstalk between adjacent waveguides. Recently, silicon nitride waveguides have been proposed for high field confinement and crosstalk reduction \cite{Zadka2018}, including for operation in the visible range \cite{ChulShin:20}.

A recent approach for implementing OPAs exploits metal \cite{Zeng2017,Lesina2020} or dielectric \cite{Komar2018,Howes2018,Li2019,Abdelsalam2019} metasurfaces. Metasurfaces are typically formed by sub-wavelength meta-atoms ({\it i.e., pixels}), thus they are naturally adapted to OPAs because the pixels can be arranged in a sub-wavelength pitch to avoid the creation of grating lobes. Im metasurfaces, the control of the phase can be achieved via electrical gating and carrier refraction effect in transparent conductive oxides (TCOs), such as indium-tin oxide (ITO) \cite{Park2017a,Huang2016,KafaieShirmanesh2018,Forouzmand2017,Lesina2020,KafaieShirmanesh2020,Forouzmand2019,Forouzmand2019a}, phase-change materials \cite{Zhu2017a,Hashemi2016,Kim2019}, liquid crystals \cite{Komar2018,Vasic2020,Wu2020,Lumotive}, the thermo-optic effect in bulk dielectrics \cite{Sun2013,Kaplan2015,Miller2020}, carrier refraction in high-index in high-index dielectric resonators \cite{Iyer2015}, ultrafast photo-carrier excitation in TCOs \cite{Alam2018a}, Fermi-level gating in graphene \cite{Sherrott2017}, or via the quantum-confined Stark effect in multi-quantum wells \cite{Lee2014c,Wu2019}. Phase control can also be achieved for nonlinear beam steering by coupling a plasmonic metasurface to a monolayer of transition metal dichalcogenides \cite{Busschaert2019}. 

Contrary to phased arrays in the microwave regime, where pixels can be ideal ({\it i.e.}, sub-wavelength, and controllable in amplitude and phase), pixels for OPAs have yet to provide such level of control.
Thus, a study of the effects of pixel limitations on beam steering quality is well-motivated and timely. All phased array technologies can be studied in terms of beam steering quality by using the phased antenna array theory, which is reviewed in this paper in Section \ref{sec_theory}. 
Beam formation depends on the array geometry -- such as the distance between pixels ({\it i.e.}, pitch), the array shape, and the array size (number of pixels) -- and the radiation properties of a single pixel. 
We describe in detail all the sources of secondary lobes in a general and technology-agnostic manner -- our results and conclusions apply to OPAs implemented in any technology. 
In Section \ref{perf_emitters}, we study the impact on beam steering performance of array properties such as pitch, shape and size for ideal pixels ({\it i.e.}, pixels that are fully controllable in amplitude and phase).
In Section \ref{nonperf_emitters}, we study how the properties of individual pixels and their imperfections affect the quality of the steered beam for perfect array conditions ({\it i.e.}, sub-wavelength pitch and ideal shape and size).
In Section \ref{sec_code}, we describe a Python code to evaluate beam steering quality in optical phased arrays, which we are making publically available \cite{PythonCode}. In Section \ref{sec_concl} we give concluding remarks.


\section{Phased array theory}\label{sec_theory}

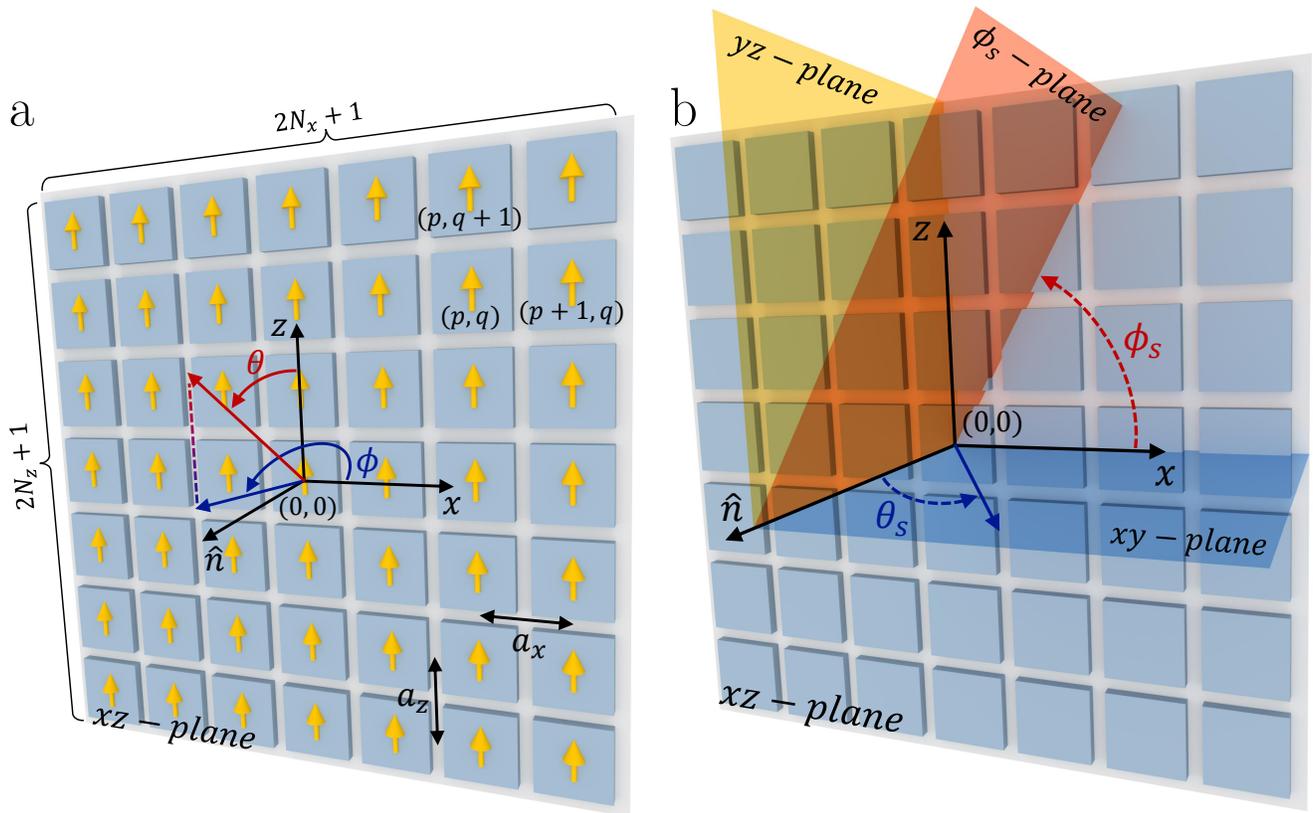
\begin{figure}[htbp]
\centering
\includegraphics[width=0.45\textwidth]{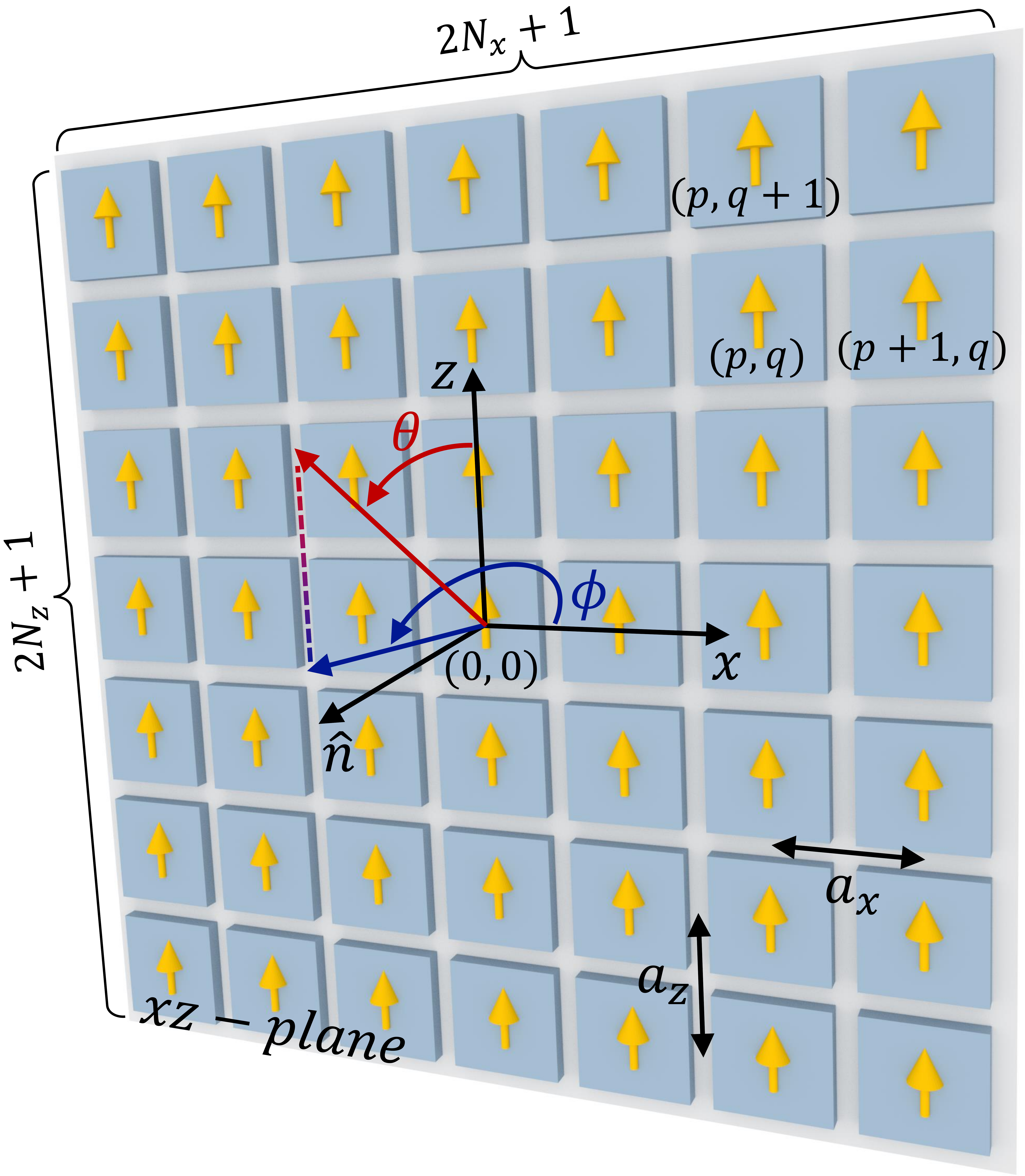}
\includegraphics[width=0.5\textwidth]{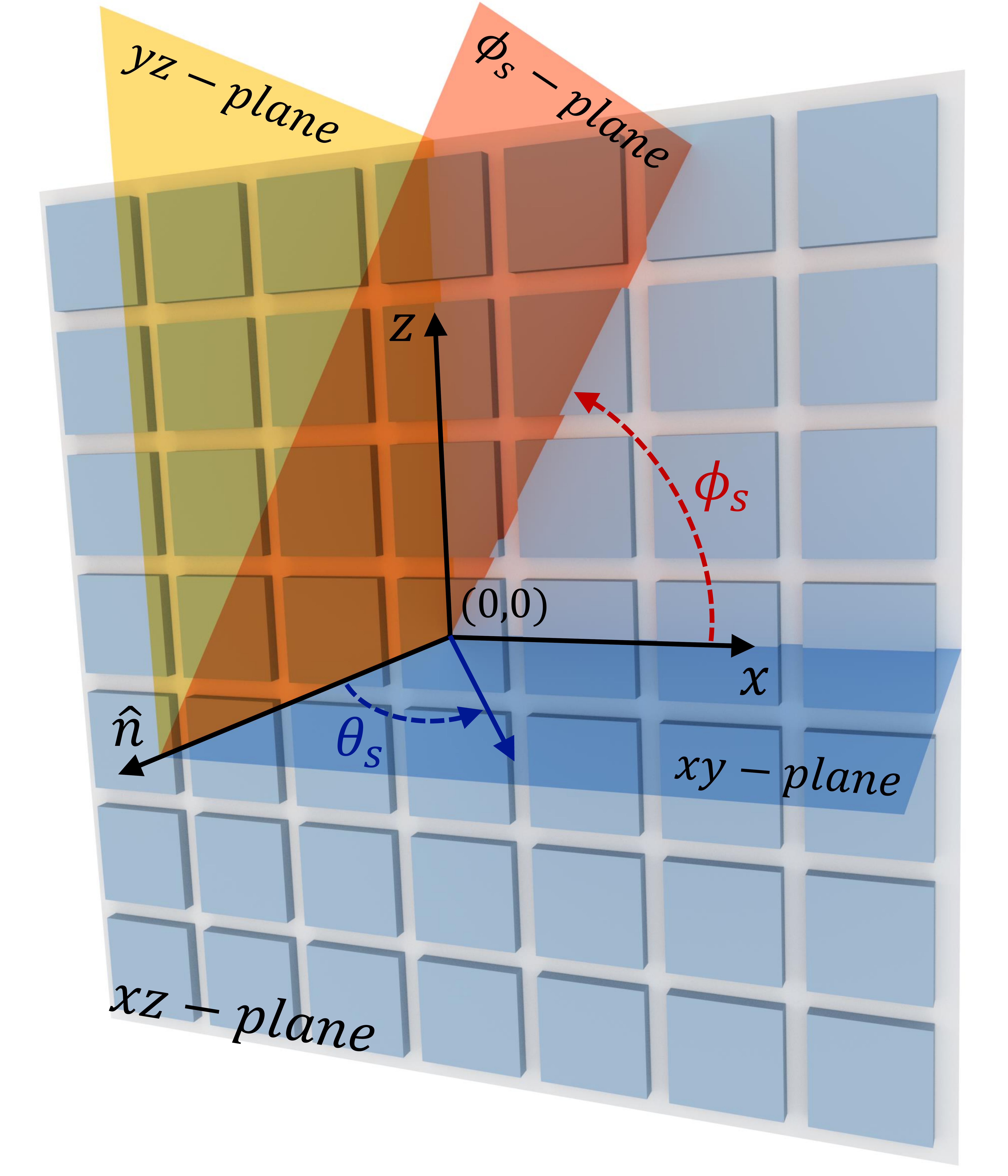}
\caption{(a) Spherical coordinates ($\theta,\phi$) to describe the radiation pattern from an array of electric dipoles, where $\theta$ is the polar angle with respect to the $z$ axis, and $\phi$ is the azimuthal angle contained in the $xy$-plane. (b) Spherical coordinates ($\theta_s,\phi_s$) to describe the beam steering from the plane of the array.}
\begin{picture}(0,0)
\put(-250,320){\huge a}
\put(0,320){\huge b}
\end{picture}
\label{fig2}
\end{figure}

In this section we review the theory for optical phased arrays.
The framework is general and valid for arrays operated either in transmission or in reflection \cite{reflectarray}, sender or receiver \cite{Fatemi2018}, or arrays composed of coherent light sources, where each pixel is driven to emit light with a specific amplitude and phase. 
We consider an array of pixels in the $xz-$plane of size $(2N_x+1)\times(2N_z+1)$, as shown in Fig. \ref{fig2}(a). We assume pixels are immersed in a dispersionless material, for example, air or glass, of refractive index $n_1$. 
The approach is valid for pixels of size $a_x\times a_z$, but for simplicity we assume a square pixel, and $a_x=a_z=a$ will be referred to as the array pitch. For generality, the pitch is normalised to the wavelength $\lambda$, with $\lambda=\lambda_0/n_1$ and $\lambda_0$ the vacuum wavelength. Thus all sizes are dimensionless, and the results are applicable over a broad range of the electromagnetic spectrum, {\it e.g.}, from radio waves to visible.

The light emission from each pixel is modelled as originating from a coherent set of radiating dipoles (yellow arrows in Fig. \ref{fig2}(a)). We describe the radiation pattern of this array by using spherical coordinates $(\theta,\phi)$, as shown in Fig. \ref{fig2}(a). Each pixel is assumed to emit a complex electric field originating from its dipole source, written in phasor notation as $E(p,q)=|E(p,q)|e^{j\psi(p,q)}$, where $(p,q)$ is the position of the pixel in the array, and $\psi$ the phase of the pixel at that location. The power radiation pattern of the array can be expressed as \cite{Barbarino,reflectarray}
\begin{equation}
|U|^2=|A|^2\cdot |F|^2,
\label{eq1}
\end{equation} 
where $A$ is the array factor,
\begin{equation}
A(\theta,\phi)=\sum_{p=-N_x}^{N_x}\sum_{q=-N_z}^{N_z}E(p,q)e^{-i k p a_x sin\theta\cos\phi}e^{-i k q a_z\cos\theta},
\label{eq_array}
\end{equation} 
and $F$ is the radiation pattern of the single pixel. Throughout the paper, we will consider dipole emitters oriented along $z$ (as sketched in Fig. \ref{fig2}(a)), with
\begin{equation}
F(\theta)=\frac{\cos(\frac{\pi}{2}\cos\theta)}{\sin\theta}.
\label{eq2}
\end{equation}

Beam steering is described using a second spherical coordinate system $(\theta_s,\phi_s)$, as shown in Fig. \ref{fig2}(b), where $\theta_s$ is the steering angle and $\phi_s$ is the angle between the steering direction and the $x$-axis.
By enforcing a phase gradient along a desired direction identified by $\phi_s$ in the $xz$-plane, beam steering is realized in the $\phi_s$-plane. 
For example, a phase gradient along $x$ ($\phi_s=0^{\circ}$) and $z$ ($\phi_s=90^{\circ}$) produces steering in the $xy$-plane and $yz$-plane, respectively. 

We use Eqs. (\ref{eq1})-(\ref{eq2}) to determine far-field array performance in the sections that follow by considering steering in the $xy$-plane -- the steering angle $\theta_s$ for this scenario is illustrated in Fig. \ref{fig2}(b). 
In the $xy$-plane, $F(\pi/2)=1$, which makes our study independent of pixel radiation pattern. However, all the results can be adapted to steering in the $yz$-plane, where $F$ depends on $\theta$ (see Eq. (\ref{eq2})). 
Furthermore, the Python code \cite{PythonCode} is general and can take into account any steering direction, array size and shape, pixel size, and pixel radiation pattern, such as a dipole emitter arbitrarily oriented in the $xz$-plane, or a more complex pixel radiation pattern (non-dipolar) input by the user.

\subsection{Beam steering and secondary lobes}

\begin{figure}[htbp]
\centering
\includegraphics[width=1\textwidth]{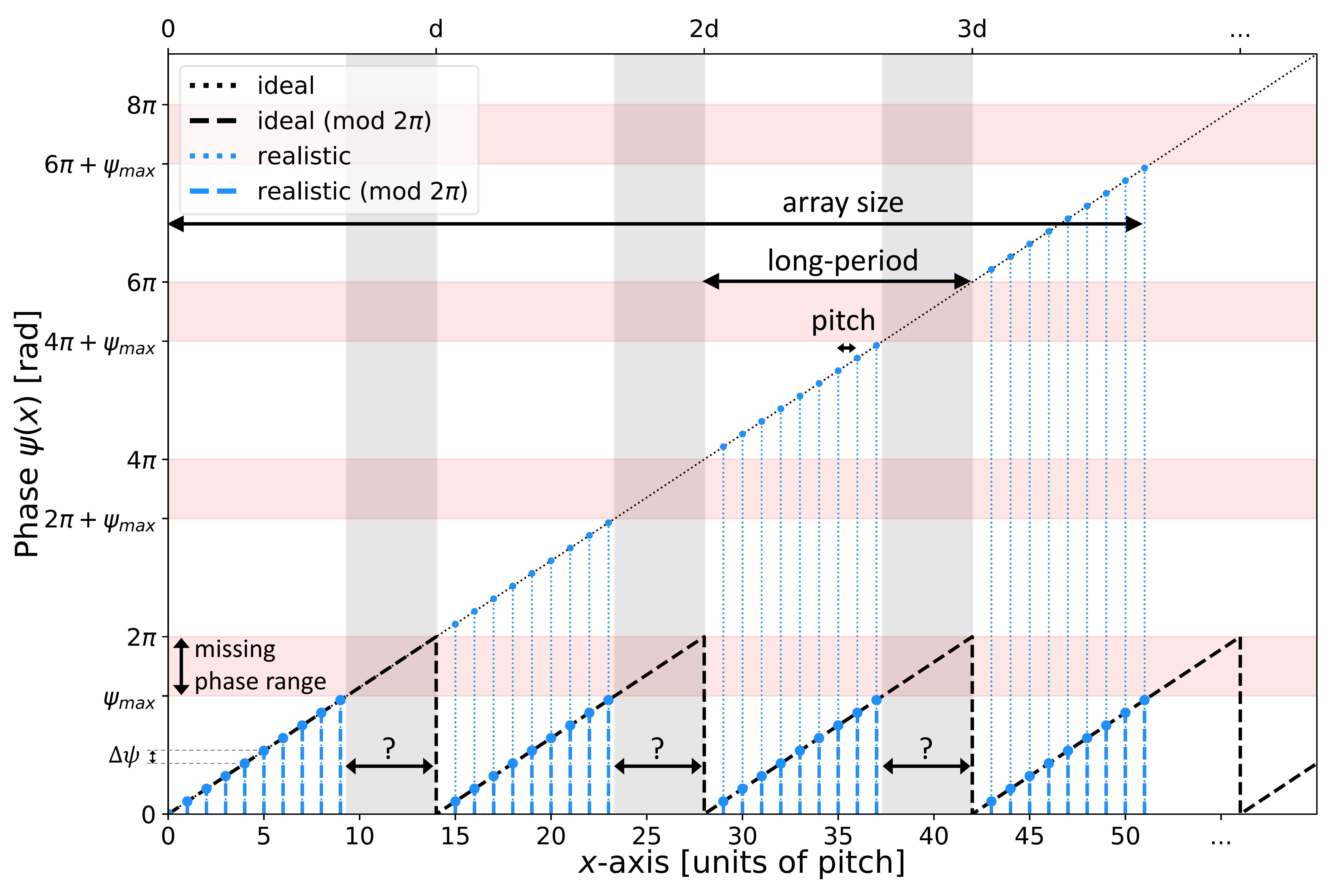}
\caption{
Ideal and realistic phase profiles as a function of pixel location. 
A finite array size leads to side lobes, a too large pitch leads to grating lobes, and a missing phase range ($\psi_{max}<2\pi$) leads to long-period grating lobes.
}
\label{fig1}
\end{figure}

To steer a beam towards a specific direction, we must produce a phase gradient across the array along that direction. For example, in order to steer in the $xy$-plane, a phase gradient along $x$ is needed  -- deriving the formulas for other steering directions is straightforward. The ideal phase distribution is continuous and is given by $\psi(x)=\psi'(x)x$, as shown in Fig. \ref{fig1} (dotted black line), where $\psi'(x)=\partial\psi(x)/\partial x$ is the phase gradient -- we also show its modulo $2\pi$ (dashed black line), that is what is implemented in practice \cite{McManamon1996}. For $\psi'(x)>0$ we have steering in the right direction, {\it i.e.}, towards a direction with $x>0$ ($\theta_s>0$), and for $\psi'(x)<0$ we have steering in the left direction, {\it i.e.}, towards a direction with $x<0$ ($\theta_s<0$). 
In fact, the term $e^{i\psi}$ introduces a delay in the time-domain signal for $\psi>0$, and $\psi'(x)>0$ means that the field is progressively delayed along $x$, thus producing steering in the same direction.
In the case of a phase profile discretized with a pitch $a$, we obtain $\psi'(x)=2\pi/d=\Delta\psi_x/a$, where $\Delta\psi_x$ is the phase difference between two adjacent pixels along $x$, {\it i.e.}, $\Delta\psi_x=\psi(p+1,q)-\psi(p,q)$, and $d$ is the sawtooth period or ``long-period'', that is large compared to the pitch, as illustrated in Fig. \ref{fig1}. Without loss of generality, we simply indicate the phase difference between adjacent pixels as $\Delta\psi$. 
The steering angle $\theta_s$ can be found from the generalized law of reflection \cite{Yu2011}:
\begin{equation}
\sin\theta_s= \frac{\psi'(x)}{k} =\frac{\lambda}{d} = \frac{\Delta\psi}{a}\frac{1}{k}, 
\label{eq1-0}
\end{equation}
where $k=2\pi n_1/\lambda_0$ is the wavenumber in the medium in which the steered beam propagates.

An ideal phase gradient for steering a plane wave requires an infinite array of infinitely small pixels with fully controllable phase and amplitude. However, this is not possible in reality, and a non-ideal phase gradient produces secondary lobes that are undesirable in LIDAR applications because they may cause interfering signal returns. 
A finite array of size smaller than the incident beam produces side lobes (similarly to an aperture); such a finite size is indicated in Fig. \ref{fig1}. A realistic phase gradient is also discretized because a pixel pitch is not infinitely small, as indicated in Fig. \ref{fig1}. However, if the pixel pitch is too large, it results in side lobes due to aliasing, in this case termed grating lobes, which may have similar strength as the main lobe.
Most pixels proposed to date for optical phased arrays are non-ideal, {\it i.e.}, they exhibit a limited phase range (the maximum phase achievable $\psi_{max}$ is less than $2\pi$), as well as a pixel amplitude that varies with its phase. 
Such non-ideal pixels introduce imperfections in the phase gradient with periodicity equal to the long-period, as sketched in Fig. \ref{fig1}, where the red horizontal band highlights the missing phase range, and the vertical grey bands indicate the pixel locations where a compensation strategy for the missing phases is needed. These long-period imperfections produce another set of side lobes, that we term ``long-period grating lobes'' (LPGLs).

In the following sections, we describe three types of secondary lobes, {\it i.e.}, side lobes, grating lobes and LPGLs, and we discuss strategies to minimize them.



\section{Effect of array limitations}\label{perf_emitters} 
In this section, we consider ideal pixels which are controllable in phase over a $2\pi$ range, and we focus on how the properties of the array, such as shape and size, and pixel pitch affect the beam steering performance. 
We consider square arrays of size $N\times N$, assuming a uniform amplitude from each pixel ({\it i.e.}, $|E(p,q)|=1$) to which we apply windowing strategies. 

\subsection{Array size and shape}\label{sec_window} 
\begin{figure}[htbp]
\centering
\includegraphics[width=1\textwidth]{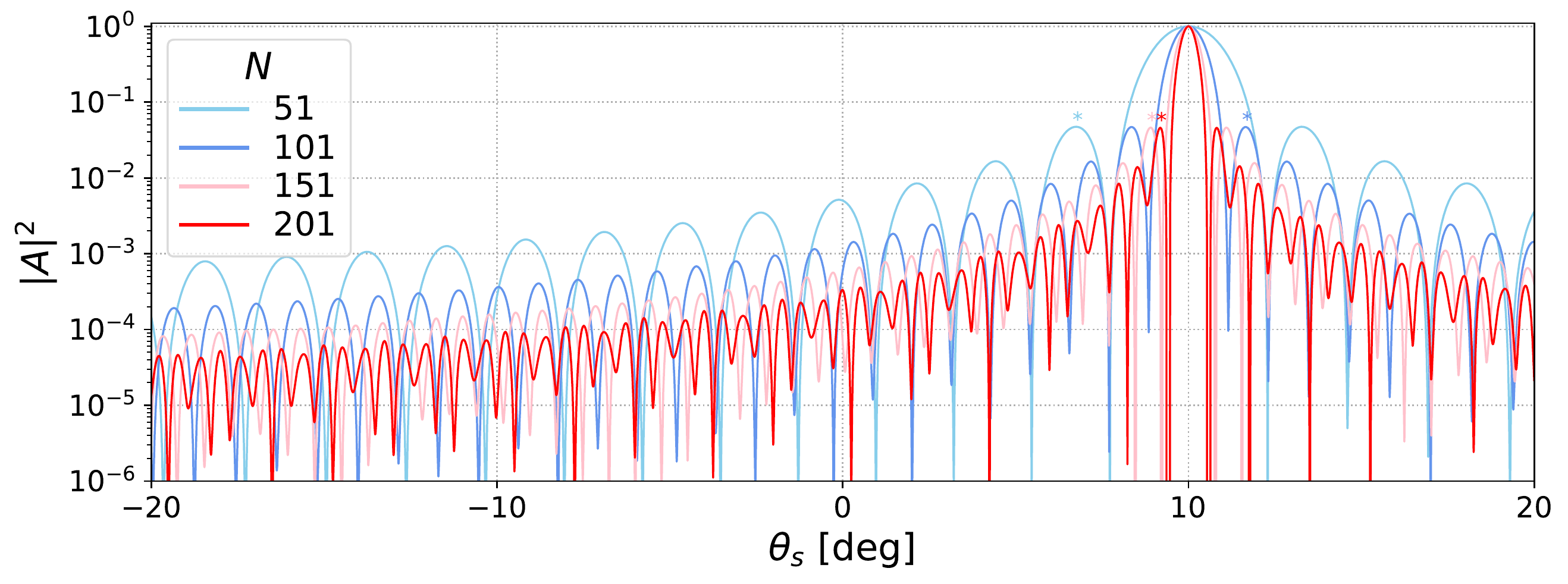}
\includegraphics[width=1\textwidth]{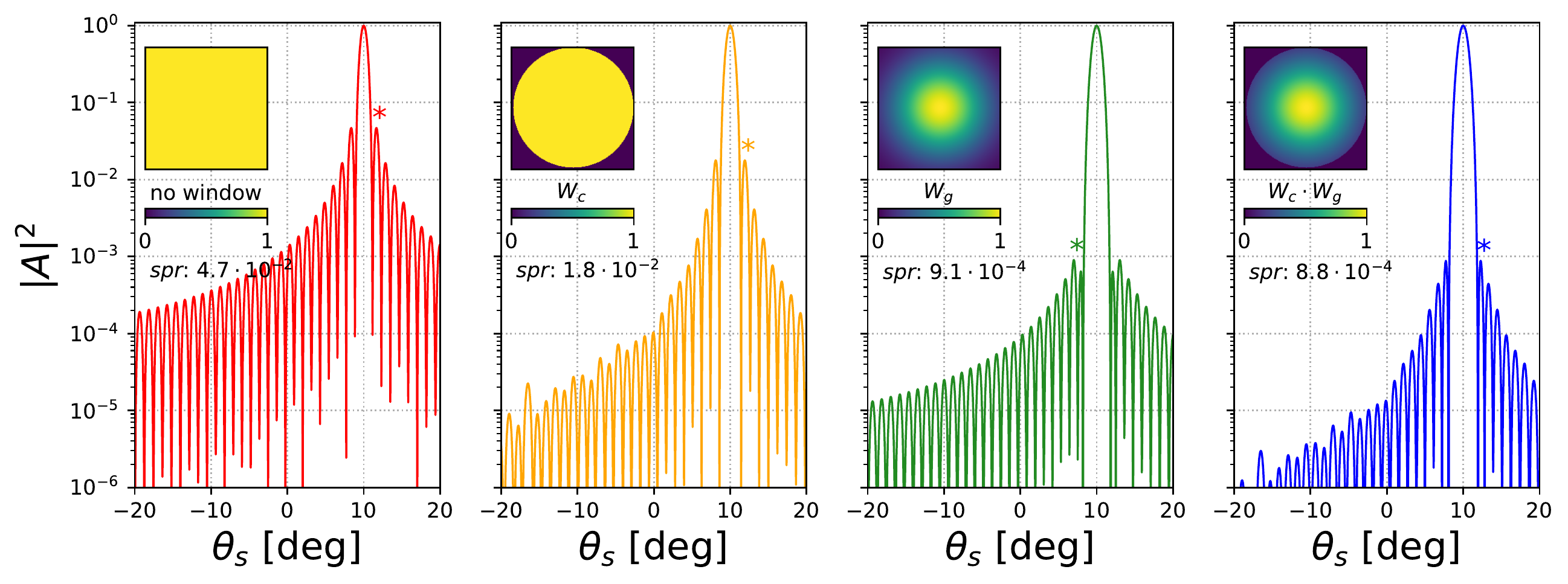}
\caption{Radiation pattern for a steering angle of $\theta_s=10^{\circ}$ for (a) different $N$, and for the following windows: (b) no window, (c) circular window, (d) Gaussian window, (e) circular and Gaussian windows.}
\begin{picture}(0,0)
\put(-250,430){\huge a}
\put(-250,230){\huge b}
\put(-110,230){\huge c}
\put(15,230){\huge d}
\put(140,230){\huge e}
\end{picture}
\label{fig3}
\end{figure}

It is knows that increasing the size of the array reduces the distance between nulls and sharpens the main lobe, thus reducing the beam width and making the radiation pattern more directive. A large array is also important for increasing resolution, as the number of phase shifters is approximately the same as the number of spots that can be resolved in the far-field \cite{Wang2019}.
Fig. \ref{fig3}(a) shows the radiation patterns for different values of $N$, with pitch $a=0.5\lambda$ and steering angle $\theta_s=10^{\circ}$. These computations model either a finite array of coherent emitters, or an infinite plane wave incident on a finite reflect- or transmittarray. 
The side lobes are caused, effectively, by diffraction from a square aperture since the array is square and finite. We also observe that the level of the highest side lobe (identified by a star ``*'') is nearly constant for varying $N$, and this applies also to the second highest, and so on. 

The side lobes in Fig. \ref{fig3}(a) are rather high at almost $10^{-1}$ of the main lobe, which may be too high for LIDAR applications. We now consider different array illumination patterns and shapes, also known as windowing, to improve on this performance.
Windowing is useful for beam steering at the transmitter or at the receiver (or both). At the transmitter, windowing may be used to improve the radiation diagram when steering a Gaussian beam that is larger than the array area. At the receiver, the illumination will be a plane wave, so windowing is essential to reducing the side lobe levels before detection. Here, we describe the effect of circular and Gaussian (apodization) windowing.
A circular window can be implemented as a circular aperture stop placed in front of the array or by realising a circular array rather than a square one. A Gaussian window can be implemented as a graduated neutral density filter placed in front of the array or by using a Gaussian beam as the incident wave instead of a plane wave (passive apodization \cite{Sun2014}). Such circular and Gaussian windows can be combined straightforwardly.    
Apodization can also be realized by actively modulating the amplitudes of the pixels across the array \cite{Zhang2019}. However, controlling the amplitude via light attenuation decreases the power efficiency, and algorithms such as those used in holography ({\it e.g.}, Gerchberg-Saxton) can be used to steer a beam only by tuning the pixel phase across the array \cite{Zhou2019}.

To facilitate comparisons with different windowing functions, we re-plot the radiation pattern of Fig. \ref{fig3}(a) in Fig. \ref{fig3}(b) for the case $N=101$. We then use windowing, which is implemented by multiplying the field $E(p,q)$ at each pixel by a circular window,
\begin{equation}
W_c(p,q)=\begin{cases}
    1, & \text{if $\sqrt{p^2+q^2}\leq N_x$}\\
    0, & \text{otherwise}
  \end{cases}
\end{equation}
a Gaussian window,
\begin{equation}
W_g(p,q)=exp\Big\{-\frac{p^2+q^2}{(\sigma\cdot N_x)^2}\Big\},
\end{equation}
or both, where $N_x=(N-1)/2$.
Note that windowing only affects the field amplitude at each pixel, and not the phase.
We quantify the beam steering quality by calculating the sidelobe-to-peak ratio ($spr$), {\it i.e.}, the ratio between the intensity of the highest side lobe (identified by a star ``*'') and that of the main lobe.
In Fig. \ref{fig3}(c) we show the case where a circular window of diameter equal to the size of the square array was applied. 
We note a lower $spr$ relative to the square case, given that diffraction by a circular aperture is comparatively weaker. If we apply a Gaussian window with $\sigma=0.75$ to the two cases of Figs. \ref{fig3}(b) and \ref{fig3}(c), we obtain the radiation patterns in Figs. \ref{fig3}(d) and \ref{fig3}(e), and note that the case where $W_c$ and $W_g$ are combined provides the best result in terms of side lobe level ($spr$ lower than $10^{-3}$). All the radiation patterns in this paper are normalized with respect to the total power emitted by the array, obtained by squaring the sum of all the pixel amplitudes. Without this normalization, for example, the level of the lobes in Fig. \ref{fig3}(b) would be higher than in Fig. \ref{fig3}(c), because the total power emitted by a square array completely illuminated is higher than that of an array with a non-uniform illumination. 

In the remainder of the paper, we assume operation of the phased array at the transmitter, and we use combined circular and Gaussian windows. Also, we will assume $N=201$ to reduce the beam width, and $\sigma=0.5$ to further reduce the side lobe level.



\subsection{Pitch and field-of-view}\label{sec_pitch}
 
The pitch of an optical phased array is often large, due to the size of the pixel, as in phased VCSELs, or to avoid mutual coupling between pixels, such as in waveguide arrays. 
A large pitch produces grating lobes, and since they may have similar amplitudes as the main beam, they are undesirable in LIDAR applications. In particular, the well known condition $a< 0.5\lambda$, that ensures steering up to $\theta_{s,max}=90^{\circ}$ without grating lobes, may be difficult to achieve for a particular technology. 
Strategies have recently been reported to suppress grating lobes, such as by engineering a sub-wavelength pitch \cite{Kossey2018,Zhang2019}, adopting a non-uniform spacing between emitters \cite{Kwong2011,Hosseini2009,Hutchison2016,Fatemi2019}, via design of a random array \cite{Sayyah2015}, and by suppressing the inter-channel coupling or crosstalk \cite{Kwong2014,Xu2019}. 

In this section, we review how the pitch affects the emergence and growth of grating lobes, and thus the FoV. 
For steering at an angle $\theta_s$, the grating diffraction equation is
\begin{equation}
\sin\theta_s^{(m)}=\sin\theta_s +m\frac{\lambda}{a},
\label{eq_grating_pitch}
\end{equation}
where $m$ is the order of the grating lobe (positive or negative integer). 
In practical applications, the FoV is usually symmetric around $\theta_s=0^{\circ}$ and defined as FoV$=\pm\theta_{s,max}$, where $\theta_{s,max}$ is the desired maximum steering angle such that there are not grating lobes in the FoV. 
However, certain applications may benefit of an asymmetric FoV. Furthermore, the FoV can be limited in the case the radiation pattern of the single pixel is narrow, such as in VCSEL pixels.
The condition $a<a_{max}$, where the upper limit for $a$ is
\begin{equation}
a_{max}=\frac{\lambda}{2|\sin\theta_{s,max}|},
\label{eq9}
\end{equation}
ensures that only the main lobe exists within the FoV. 
This equation shows that when $\theta_{s,max}$ ({\it i.e.}, FoV) decreases, $a_{max}$ increases. Thus, in applications where a smaller FoV is needed, the requirements for $a<a_{max}$ are relaxed. For example, for a desired $\theta_{s,max}=5^{\circ}$, $a_{max}=5.74\lambda$ is required; for $\theta_{s,max}=10^{\circ}$, $20^{\circ}$, $30^{\circ}$ and $50^{\circ}$, then $a_{max}$ is $2.88\lambda$, $1.46\lambda$, $\lambda$ and $0.65\lambda$, respectively. 

\begin{figure}[htbp]
\centering
\includegraphics[width=1\textwidth]{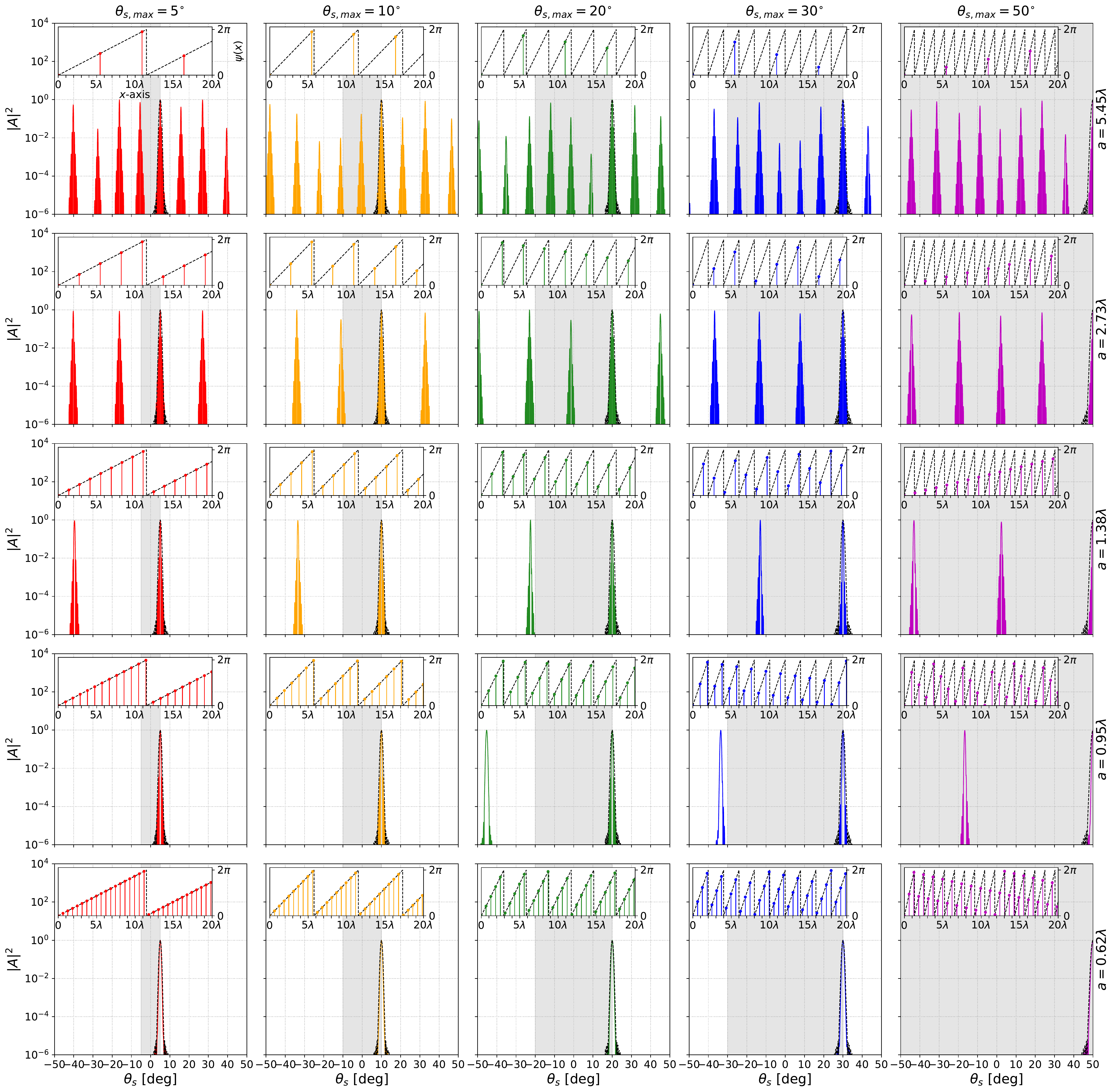}
\caption{Radiation diagrams for different values of $a<a_{max}$ and $\theta_{s,max}$ (the FoV is highlighted by a grey band in each sub-figure). 
}
\label{fig4}
\end{figure}

This is confirmed in Fig. \ref{fig4} by computing the radiation diagram for different cases of $a$ and $\theta_{s,max}$. In each sub-figure, the FoV is highlighted by a grey band and we also show the ideal steered beam for $a=0.5\lambda$ (black dashed line).
In the insets, we show the ideal phase profile (black dashed line) and the realistic one (coloured line) over a portion of the array of length $20\lambda$ for illustration convenience. 
For $\theta_{s,max}=5^{\circ}$, all $a$ values are smaller than $5.74\lambda$, and we do not see grating lobes in the FoV. For $\theta_{s,max}=10^{\circ}$, the case $a=5.44\lambda$ does not satisfy the condition $a<a_{max}$, and we see a grating lobe in the FoV. For $\theta_{s,max}=50^{\circ}$, only the case $a=0.62\lambda$ satisfies the condition $a<a_{max}$, and we see grating lobes in the FoV for all the other $a$ values. 



\section{Effect of pixel limitations}\label{nonperf_emitters}
Ideally, beam steering requires full control of all pixels in the array, that is, full control over $E(p,q)$. However, in practice, optical pixels have limitations that preclude such full control. For example, they may not provide a full $2\pi$ phase range, or they may not provide control of the amplitude (or even maintain a constant amplitude) over the phase range. Such limitations will lead to other types of secondary lobes. In what follows, we investigate the effect of several common pixel limitations, while assuming the best array configuration possible, {\it i.e.}, circular and Gaussian ($\sigma=0.5$) windows, and $a=0.5\lambda$. 


\subsection{Pixels with limited phase range}\label{sec_phase}

\begin{figure}[htbp]
\centering
\includegraphics[width=0.8\textwidth]{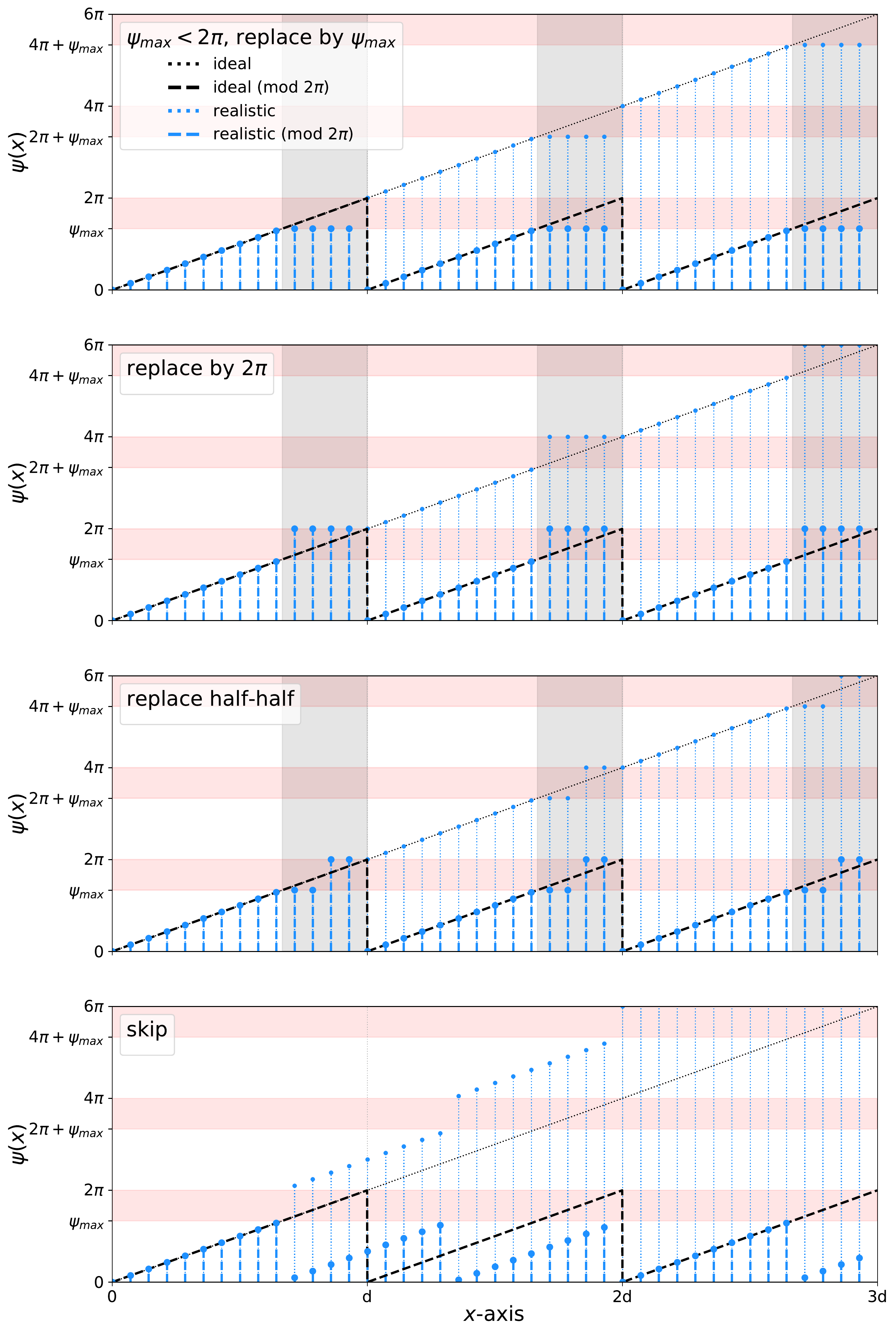}
\caption{Compensation strategies when $\psi_{max}<2\pi$: (a) replace by $\psi_{max}$, (b) replace by $2\pi$, (c) replace half-half, (d) skip.}
\begin{picture}(0,0)
\put(-230,660){\huge a}
\put(-230,500){\huge b}
\put(-230,345){\huge c}
\put(-230,190){\huge d}
\end{picture}
\label{fig5}
\end{figure}

\begin{figure}[htbp]
\centering
\includegraphics[width=1\textwidth]{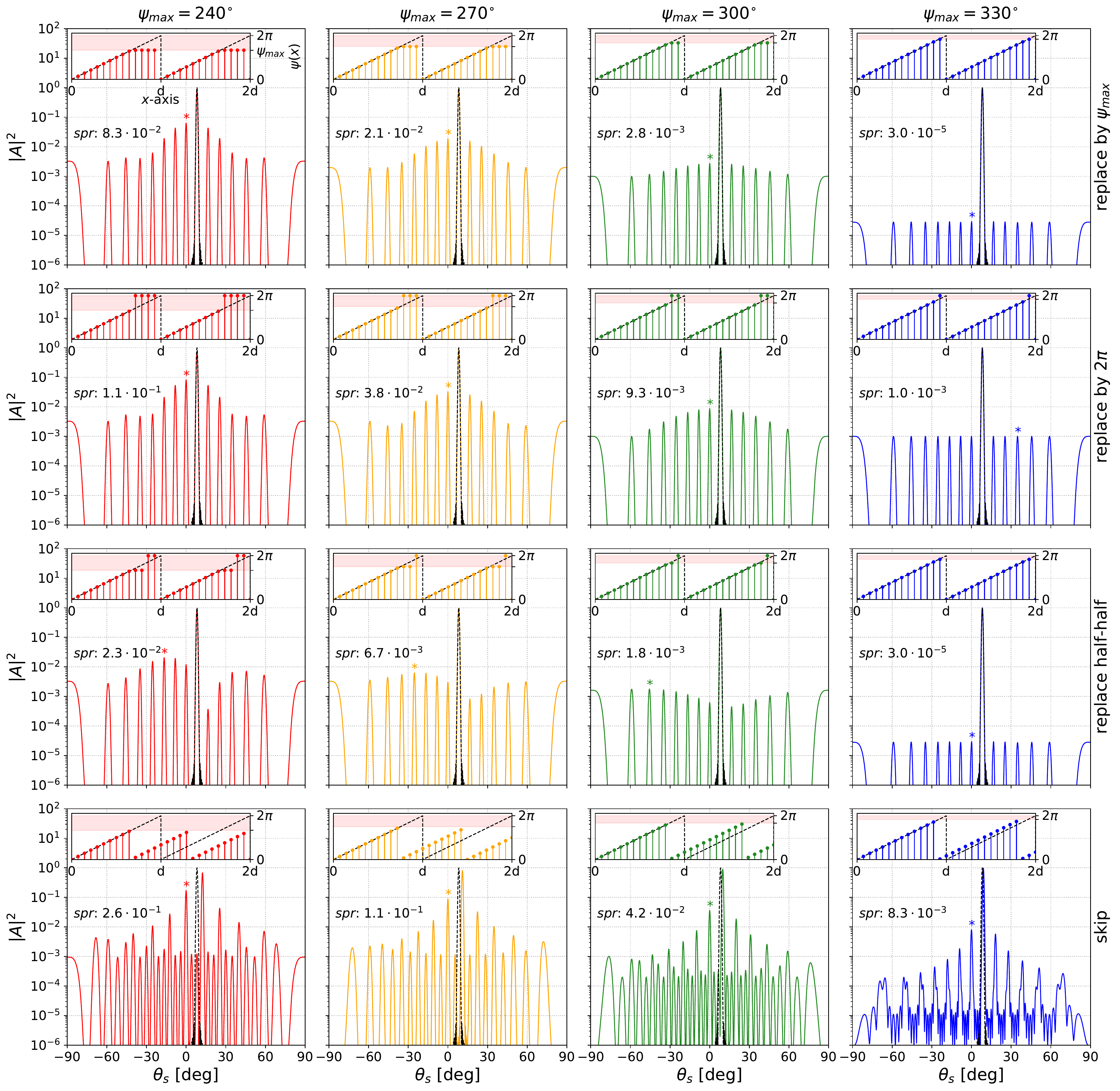}
\caption{Long-period grating lobes due to different phase compensation strategies (illustrated in the insets), for varying $\psi_{max}$ and for steering at $\theta_s = 8.21^{\circ}$ ($M=14$, $a=0.5\lambda$).}
\label{fig6}
\end{figure}

Here, we study how a limited pixel phase range (max$\{\angle E(p,q)\} = \psi_{max} < 2\pi$) affects the main lobe and the characteristics of secondary lobes. As illustrated in Fig. \ref{fig1}, beam steering requires that a phase gradient be created along the steering direction. In practical applications, a realistic phase gradient is implemented by using pixels of finite size as shown in Fig \ref{fig1}. Also, what is implemented in applications is the modulo $2\pi$ of the continuous phase, resulting in a sawtooth phase profile of period $d$,
\begin{equation}
d =\frac{2\pi}{\psi'}= \frac{2\pi}{k\cdot\sin\theta_s}=\frac{\lambda}{\sin\theta_s}=M\cdot a,
\label{eq_d}
\end{equation}
where the second term is obtained by using Eq. (\ref{eq1-0}), and $M$ is the number of pixels per sawtooth period (which is also the number of pixels contained within $2\pi$),
\begin{equation}
M=\frac{d}{a}=\frac{2\pi}{\Delta\psi}.
\end{equation}

In previous literature, $M$ has been considered as an integer 
\cite{McManamon1996,McManamon2009,Vasic2020}. However, 
$M$ can be any real number, and in this section we discuss both cases. 
We can rewrite Eq. (\ref{eq_d}) as
\begin{equation}
\sin\theta_s= \Big(\frac{a}{\lambda}\cdot M\Big)^{-1}.
\label{eq_angles_M}
\end{equation}
Examining Eq. (\ref{eq_angles_M}), we notice that $a/\lambda$ depends on the technology, and $M$ determines the steering angles that can be obtained. Thus, only considering integer $M$ would lead to a discrete set of steering angles, rather than a continuous beam steering.

When $\psi_{max}<2\pi$, the discretized sawtooth phase profile cannot be fully achieved because some phase values are not available, as illustrated in Fig. \ref{fig1}. This results in a sawtooth with defects. These defects follow the periodicity $d$ of the sawtooth, which we also call ``long-period'' because greater than the pitch. The secondary lobes due to the defects in the sawtooth are named ``long-period grating lobes'' (LPGLs). Writing the grating diffraction equation in terms of the long-period $d$, we obtain:
\begin{equation}
\sin\theta_s^{(l)}=\sin\theta_s+\frac{l'\lambda}{d} = \frac{\lambda}{d} +\frac{l'\lambda}{d} =l\sin\theta_s,
\label{eq_grating_max_phase}
\end{equation}
where we used Eq. (\ref{eq_d}), and where $l=1+l'$ is the diffractive order of a LPGL. The maximum order for an LPGL, $l_{max}$, emerges when $\sin\theta_s^{(l_{max})} = 1$, giving
\begin{equation}
l_{max}=\mathrm{int}\Big(\frac{1}{|\sin\theta_s|}\Big).
\label{eqlmax}
\end{equation}
The number of LPGLs is $2l_{max}$, which increases as the steering angle $\theta_s$ decreases. Also, $l$ varies between $-l_{max}$ and $l_{max}$, and $l=1$ identifies the main lobe at $\theta_s$, that is not counted as an LPGL.
Note that LPGLs caused by missing phase values are more tightly spaced than grating lobes produced by a too-large pitch. Thus, they cannot generally be avoided by restricting the desired maximum steering angle, {\it i.e.}, the FoV.

The number $M$ is important also in the case of pixels with limited phase range. In fact, ideal steering (with no LPGLs) can be obtained even when the pixel phase does not cover the $2\pi$ range. Given an integer $M$, ideal steering at $\theta_s$ is possible if $\psi_{max}\geq 2\pi\cdot M/(M+1)$ based on Eq. (\ref{eq_angles_M}).
For example, for $a=0.5\lambda$ and $M=14$ ($M=4$), we can steer at $\theta_s=8.21^{\circ}$ ($\theta_s=30^{\circ}$) without LPGLs given that $\psi_{max}\geq 336^{\circ}$ ($\psi_{max}\geq 288^{\circ}$).

As already discussed in Fig. \ref{fig1}(b), a limited phase range requires compensation strategies for the missing phase values. 
We investigate four compensation strategies, as shown in Fig. \ref{fig5}: (i) replacing the missing phases with $\psi_{max}$ (``replace by $\psi_{max}$''); (ii) replacing the missing phases with $2\pi$ (``replace by $2\pi$''); (iii) replacing one half of the missing phases with $\psi_{max}$ and the other half with $2\pi$ (``replace half-half''); and (iv) skipping the missing phases and restart the sawtooth immediately to $\psi=0^{\circ}$. 
Optimization algorithms and artificial intelligence can also be used to find non-intuitive phase compensations.

We plot in Fig. \ref{fig6} the computed radiation patterns that result from our compensation strategies. We also add in each case the radiation pattern produced by the ideal phase gradient as the black dashed curves. The strategies already introduced in Fig. \ref{fig5}, are illustrated in the insets of each sub-figure of Fig. \ref{fig6},
where we consider different values of $\psi_{max}$, {\it i.e.}, $\psi_{max}=240^{\circ}$, $270^{\circ}$, $300^{\circ}$, and $330^{\circ}$, and we use an integer number of pixels to discretize $d$, {\it i.e.}, $M=14$, thus resulting in $\theta_s = 8.21^{\circ}$.
This steering angle produces $2l_{max}=14$ LPGLs, following Eq. (\ref{eqlmax}). We fix $|E(p,q)|=1$ (before windowing), so that the secondary lobes are only due to the limited pixel phase range. 

For each sub-figure in Fig. \ref{fig6}, we also report the $spr$ (the largest LPGL is identified by ``*'').
We see that the $spr$ decreases with increasing $\psi_{max}$ for all compensation strategies, and that the ``replace half-half'' strategy is best. This is not surprising, as it best approximates the phase gradient in the range where the phase values are missing. The ``replace by $\psi_{max}$'' strategy gives smaller $spr$ than ``replace by $2\pi$'', while the ``skip'' strategy gives the worst result. However, skip also alters the phase gradient and thus the steering angle, as noted by comparison of the skip main lobe with the ideal main lobe (black dashed). Skip also gives a larger number of LPGLs since the sawtooth period is modified, $d^{(skip)}=\psi_{max}/\Delta\psi$, and $M$ is now non-integer.
Importantly, these results show that it is possible to realize high-quality beam steering even for $\psi_{max}<2\pi$, as the LPGLs can be two orders of magnitude smaller than the main lobe, which is adequate for many applications.


\begin{figure}[htbp]
\centering
\includegraphics[width=1\textwidth]{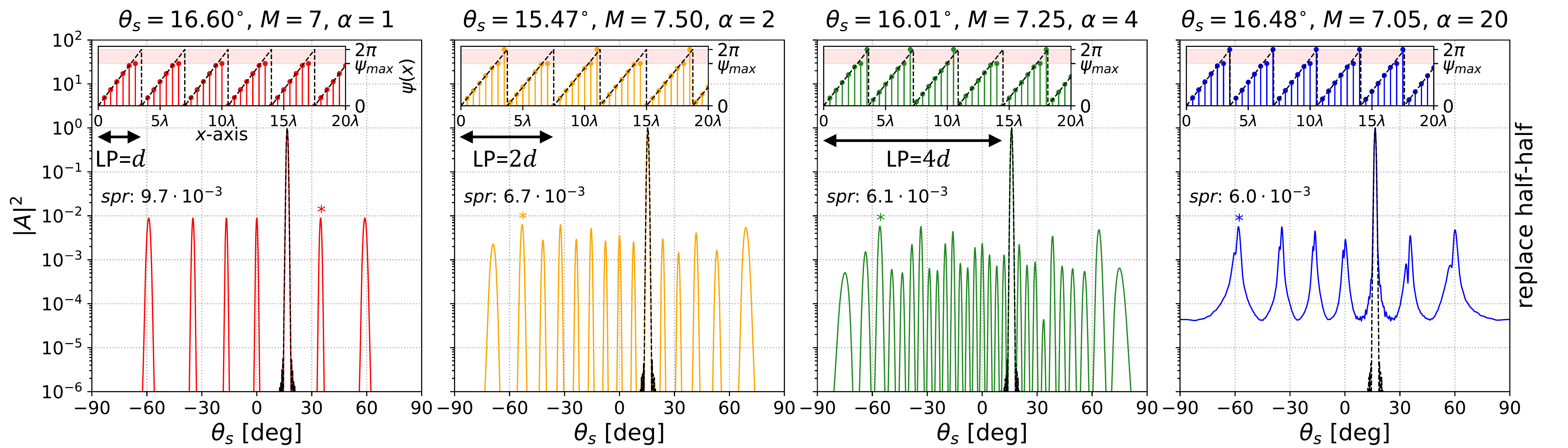} 
\caption{Long-period grating lobes for different values of $M$ (integer and non-integer), $\psi_{max}=270^{\circ}$ and ``replace half-half'' strategy. From left to right we have 6, 14, 28, and 140 LPGLs, respectively.}
\label{fig7}
\end{figure}

We have considered so far $M$ as an integer.
Thus, the only long-period introduced by the missing phases is $d$. In general, $M$ can assume any real value, and this is also valid for $d$ and $\theta_s$ (see Eq. (\ref{eq_d})).
This modifies the long-period in the phase gradient to $\alpha d$, with $\alpha$ such that $\alpha M$ is an integer. For example, $M=7.5$, $7.25$ and $7.05$ correspond to $\alpha=2$, $4$ and $20$, meaning that the long-period is $2d$, $4d$ and $20d$; these cases are shown in Fig. \ref{fig7}, along with the case $M=7$, for $\psi_{max}=270^{\circ}$, using the replace half-half strategy. In the case $\alpha=20$, the number of LPGLs is so high that they cannot be distinguished.
For $M$ non-integer, the orders and angles of the LPGLs are obtained by introducing the long-period $\alpha d$ in Eqs. (\ref{eq_grating_max_phase}) and (\ref{eqlmax}):
\begin{equation}
\sin\theta_s^{(l)} = \sin\theta_s+\frac{l'\lambda}{\alpha d} = \frac{l}{\alpha}\sin\theta_s, 
\label{eq_grat_alpha}
\end{equation}
where $l=\alpha+l'$ and the maximum LPGL order becomes 
\begin{equation}
l_{max}=\mathrm{int}\Big(\frac{\alpha}{|\sin\theta_s|}\Big). 
\label{eq_max_alpha}
\end{equation}
Although a non-integer $M$ increases the number of LPGLs, we notice that the lobes of order $l$ such that $l/\alpha$ is an integer in Eq. (\ref{eq_grat_alpha}) remain higher in amplitude -- these lobes are identified by Eq. (\ref{eq_grating_max_phase}). We conclude that considering only the periodicity $d$ for the evaluation of the LPGLs and for the calculation of the $spr$ is accurate in most cases. In Fig. \ref{fig7}, we also notice that the $spr$ is higher for an integer $M$ with respect to the non-integer cases due to the fact that the total radiated power is concentrated in fewer lobes, {\it i.e.}, choosing an integer $M$ is a worst case scenario. Thus, in real applications, where the steering angle can assume any value and $M$ is most likely non-integer, the $spr$ can be lower than what predicted for integer $M$ values.

\subsection{Pixels with varying amplitude}\label{sec_ampl}

Optical pixels often produce a response where the amplitude of the field cannot be controlled (or maintained constant) independently from the phase. Thus, altering the phase often results in perturbing the amplitude which is generally undesirable. Here, we investigate the effects of varying amplitude by assuming $\psi_{max}=2\pi$. 

We consider a perturbed electric field across the array by adding a perturbing function, $f_p(x)$, to $|E(x)|=1$ before windowing, {\it i.e.}, $|\tilde E(x)|=1+f_p(x)$. 
We choose the perturbing function to be a sum of an offset constant $A$ and a sinusoidal function with amplitude $B$ and period $d/P_d$, 
\begin{equation}
f_p(x)=A+B\cdot\sin\Big(2\pi \frac{x}{d/P_d}\Big),
\end{equation}
where $P_d$ controls the shape of the perturbation. In fact, $P_d$ represents the number of cycles of the sinusoidal function within $d$, {\it i.e.}, 
\begin{equation}
2\pi \frac{x}{d/P_d}=P_d\cdot\psi'(x)x=P_d\cdot\psi(x),
\end{equation}
revealing the dependence of the sinusoidal perturbation on the pixel phase $\psi(x)$.
For a perturbation containing full sinusoidal cycles ($P_d\geq1$), we have $A=0$ because the sinusoid is symmetric about unity. For $P_d\leq 0.5$, we need to use $A \neq 0$ that, by vertically translating the sinusoid, guarantees that the maximum and minimum of $|\tilde E(z)|$ are equidistant from unity, {\it i.e.}, $(|\tilde E(x)|_{max}+|\tilde E(x)|_{min})/2=1$. Varying $P_d$, while tuning $B$ and $A$, allows us to obtain different perturbation shapes of the same strength $|\tilde E|_{var}$, which we define as
\begin{equation}
|\tilde E|_{var}=\frac{|\tilde E(x)|_{max}-|\tilde E(x)|_{min}}{|\tilde E(x)|_{max}+|\tilde E(x)|_{min}}
\end{equation}
and is expressed in what follows as a percentage. 


\begin{figure}[htbp]
\centering. 						              
\includegraphics[width=1\textwidth]{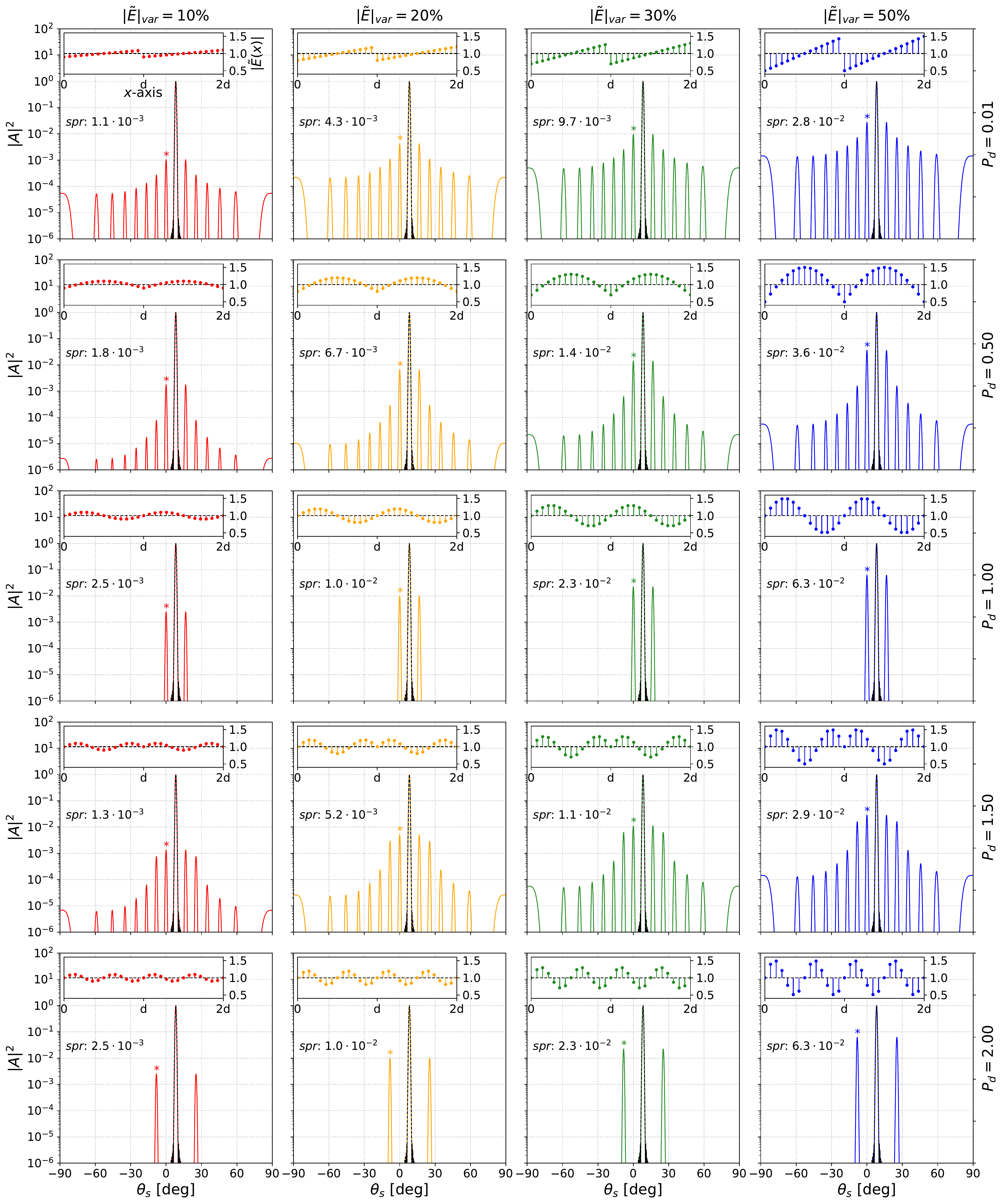}
\caption{Long-period grating lobes due to perturbations in the pixel amplitude of varying strengths, $|\tilde E|_{var}$, and shapes, $P_d$. Same nominal steering set up at $\theta_s = 8.21^{\circ}$ as for Fig. \ref{fig6} ($M=14$, $a=0.5\lambda$).}
\label{fig8}
\end{figure}

In Fig. \ref{fig8}, we plot the far-field radiation patterns resulting from different amplitude perturbations applied over a long-period $d$ for $|\tilde E|_{var}=10\%$, $20\%$, $30\%$ and $50\%$, and $P_d=0.01$, 0.5, 1, 1.5 and 2. Also in this case, we consider $M=14$ and $\theta_s = 8.21^{\circ}$. The insets plot the ideal amplitude of $E(x)$ before windowing, {\it i.e.}, $|E(x)|=1$ (black dashed line), as well as the amplitude of the perturbed field, $|\tilde E(x)|$ (coloured curves).
Unsurprisingly, the $spr$ in Fig. \ref{fig8} increases with $|\tilde E|_{var}$ but not necessarily with $P_d$. We note for non-integer $P_d$ that the LPGLs occur at the same diffraction angles already reported in Section \ref{sec_phase} for the LPGLs due to missing pixel phases. 
For an integer $P_d$, we have less orders but they have nearly double amplitude -- given that $l=1$ is the main lobe, we have LPGLs of orders $l=1 \pm P_d$. 

\subsection{Combined effect of limited phase range and varying amplitude}\label{sec_phase+ampl}

\begin{figure}[htbp]
\centering
\includegraphics[width=0.7\textwidth]{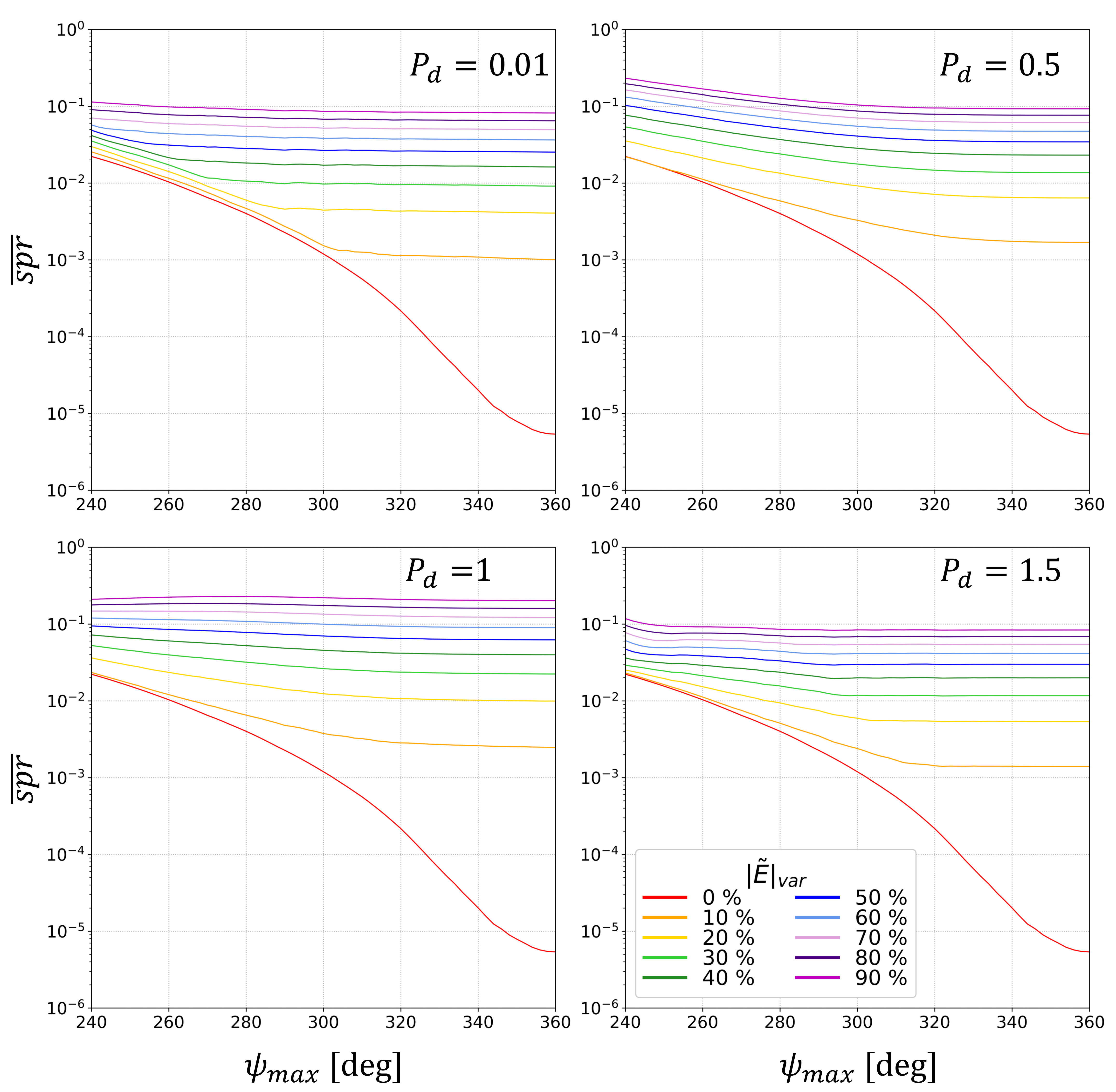}
\caption{Average $spr$ for steering angles between $1^{\circ}$ and $90^{\circ}$, for a limited pixel phase range (varying $\psi_{max}$) and a non-uniform pixel amplitude (varying $|\tilde E|_{var}$ and $P_d$).}
\label{fig9}
\end{figure}

In this section, we summarize our findings by considering a broader set of steering angles and pixel limitations. In Fig. \ref{fig9}, we plot the average $spr$ ({\it i.e.}, $\overline{spr}$) for steering angles between $1^{\circ}$ and $90^{\circ}$ in steps of $1^{\circ}$, for $a<0.5\lambda$, for different values of $P_d$ and $|\tilde E|_{var}$, and for varying $\psi_{max}$ (applying the replace half-half strategy). The case $P_d=0.01$ produces the lowest levels for the LPGLs. We also note a large change in $\overline{spr}$ as $|\tilde E|_{var}$ varies from $0\%$ to $10\%$, which reinforces the notion that the amplitude profile strongly affects the radiation pattern. For a given $|\tilde E|_{var}$, we see that the $\overline{spr}$ is constant over a range of $\psi_{max}$. This clearly separates the regime where LPGLs are dominated by the varying amplitude, from the regime where LPGLs are dominated by the limited phase range. If we set $\overline{spr} =10^{-2}$ as the maximum value that we can tolerate in practical applications, we find that the sawtooth amplitude perturbation ($P_d=0.01$) allows us to obtain this ratio even with $\psi_{max}\sim 270^{\circ}$ and $|\tilde E|_{var}\sim 30\%$, which are achievable with many pixel designs. We also note that $\psi_{max}=260^{\circ}$ represents a limit in that no beam steering with $\overline{spr}< 10^{-2}$ can be realized for $\psi_{max}< 260^{\circ}$.

\section{Python code}\label{sec_code}

We provide a Python code \cite{PythonCode} that exploits distributed computing to calculate the radiation pattern of an array with the following input parameters: 
\begin{itemize}
\item steering angle $\theta_s$, 
\item steering plane (identified by $\phi_s$), 
\item polarization direction of all dipoles, 
\item size of the array $(2N_x+1)\times (2N_z+1)$, 
\item windowing strategy (circular and/or Gaussian), 
\item pitch size $a_x\times a_z$ (both normalized with respect to $\lambda$), 
\item maximum pixel phase $\psi_{max}$, 
\item phase compensation strategy, 
\item shape of the amplitude perturbation (through the parameter $P_d$), 
\item strength of the amplitude perturbation $|\tilde E|_{var}$ (through the parameters $A$ and $B$). 
\end{itemize}

\begin{figure}[htbp]
\centering
\includegraphics[width=1\textwidth]{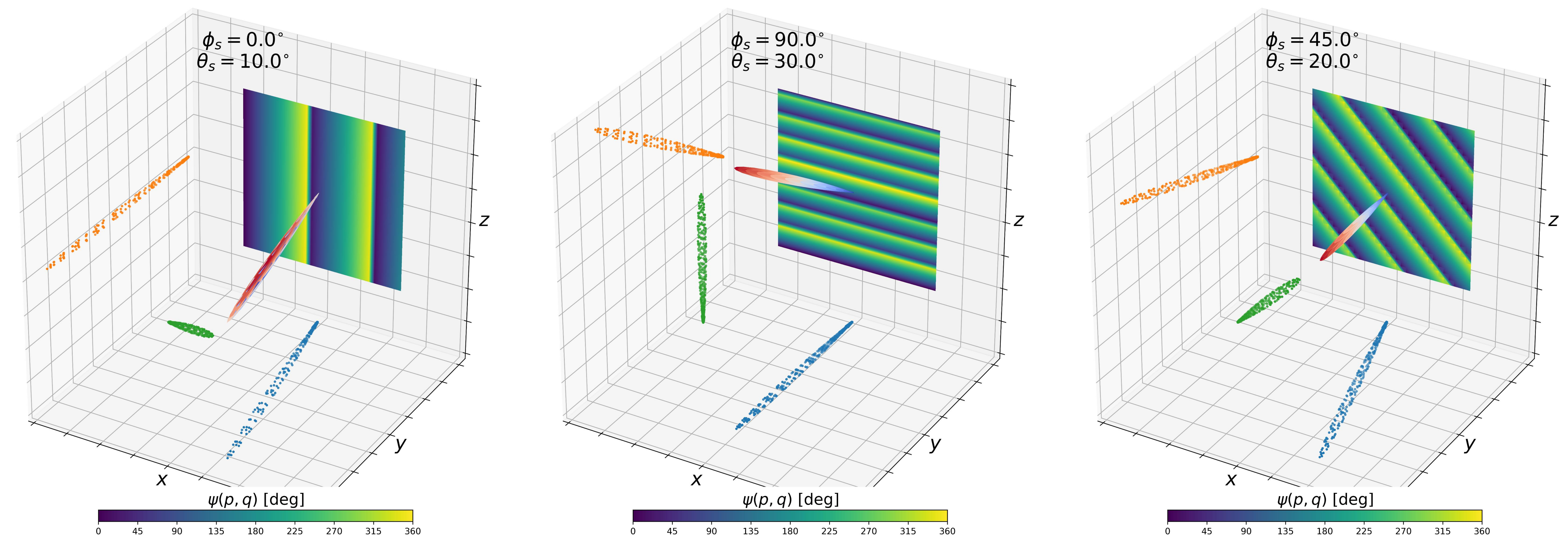}
\caption{Illustration of beam steering in 3D in the (a) $xy$-plane with $\theta_s=10^{\circ}$ \cite{Movie1}, (b) $yz$-plane with $\theta_s=30^{\circ}$ \cite{Movie2}, (c) $45^{\circ}$-plane with $\theta_s=20^{\circ}$ \cite{Movie3}.}
\begin{picture}(0,0)
\put(-260,210){\huge a}
\put(-80,210){\huge b}
\put(100,210){\huge c}

\end{picture}
\label{fig10}
\end{figure}
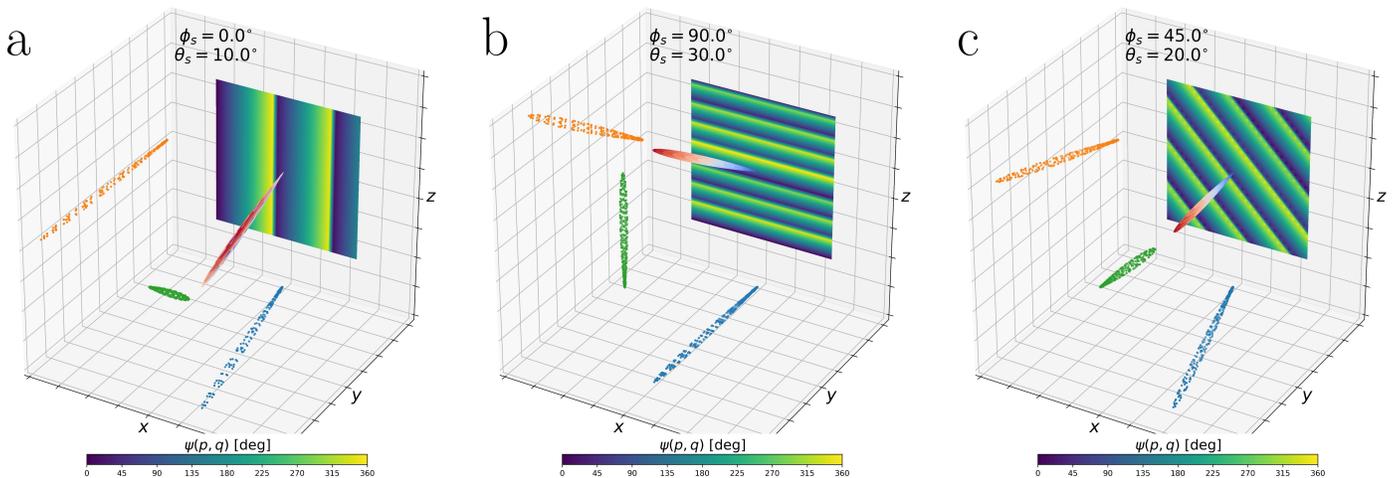

In particular, the Python code can generate 3D radiation patterns as shown in Fig. \ref{fig10} for different values of steering direction $\phi_s$ and steering angle $\theta_s$. Figures \ref{fig10}(a), \ref{fig10}(b) and \ref{fig10}(c) are extracted from Visualizations 1 \cite{Movie1}, 2 \cite{Movie2}, and 3 \cite{Movie3}, respectively. These movies show beam steering in the horizontal ($\phi_s=0^{\circ}$), vertical  ($\phi_s=90^{\circ}$) and diagonal plane ($\phi_s=45^{\circ}$), by considering $N=31$ elements without windowing. Here, the effect of the dipole radiation pattern is visible by comparing steering in the $yz$ and $xy$ planes.
The square images in each sub-figure of Fig. \ref{fig10} represent the phase profile across the array to produce beam steering. Furthermore, the 3D beam is projected in the $xy$ (blue dots), $xz$ (green dots) and $yz$ (orange dots) planes to better illustrate its orientation in space.
 
In these visualizations we have used ideal conditions, {\it i.e.}, $a<0.5\lambda$, $\psi_{max}=2\pi$, and constant amplitude, but the user can generate 3D plots in presence of pixel and array imperfections to visualize the effect of such limitations on beam steering and secondary lobes. Although throughout the paper, we have only considered steering in the horizontal direction ($xy$-plane), the Python code \cite{PythonCode} is general and allows us to steer the beam in any plane, {\it i.e.}, to change $\phi_s$ in Fig. \ref{fig2}(b). Having access to the code, the user can introduce new phase compensation strategies, amplitude profiles, and input a pixel radiation pattern that is more complex than a dipole emitter. The code also allows to save the data for the 2D radiation patterns plotted throughout the paper, which can be used for further analysis.

\section{Conclusion}\label{sec_concl}
Pixels for optical phased arrays in practice have limitations in terms of size, phase range and amplitude control. All of these characteristics affect the quality of the steered beam by reducing the amplitude of the main lobe and generating undesired secondary lobes, such as grating lobes due to a too-large pitch and long-period grating lobes due to an incomplete phase range and/or non-constant amplitude, which occur over much longer length scales than the pitch. Our study investigates the effect of many common sources of undesired lobes, as well as presents strategies to minimize them. 
We have discussed the effect of different windowing strategies on side lobes, identifying the combination of circular and Gaussian windowing as best in terms of side lobe level. 
We have shown that the pitch size can be increased and grating lobes can be avoided if the desired steering range is reduced accordingly. We have discussed the effect of pixel limitations, such as a limited phase, and discussed the performance of different strategies to compensate for the missing phase range. Furthermore, we have considered the effect of a non-uniform pixel amplitude by evaluating the performance for different amplitude perturbation shapes. 
We find that a sawtooth perturbation of the pixel amplitude allows achieving a sidelobe-to-peak ratio of $10^{-2}$ with a maximum pixel phase down to $\sim 270^{\circ}$, and a percentage variation of the pixel amplitude up to $\sim 30\%$
These values are obtainable in many pixel technologies, and this makes optical phased arrays based on those pixel technologies viable for practical applications. The attached Python code is handy to visualize/analyze in 3D and 2D how the parameters of the array affect the steered beam. The generality of our study can be applied by researchers from different communities working on optical phased arrays for beam steering for applications in LIDAR technology and smart communications.

\section*{Acknowledgments}
We acknowledge computational support from SciNet and Compute Canada, and financial support from NSERC and Huawei Technologies Canada.

\section*{Disclosures}
The authors declare that there are no conflicts of interest related to this article.

\bibliography{Lidar_paper2}

\end{document}